\documentclass{PoS}
\usepackage{amsmath}
\usepackage{bm}
\usepackage{mathtools}
\RequirePackage{xspace}

\def\invab{\ensuremath{\mbox{\,ab}^{-1}}\xspace}
\def\Y#1S{\ensuremath{\Upsilon{(#1S)}}\xspace}
\newcommand{\gev}{\ensuremath{\mathrm{\,Ge\kern -0.1em V}}\xspace}
\title{Prospects for Rare Decays at Belle II}

\ShortTitle{Prospects for Rare Decays at Belle II}

\author{\speaker{S. Sandilya}\thanks{On behalf of the Belle II Collaboration}\\
        University of Cincinnati, Cincinnati, Ohio 45221\\
        E-mail: \email{saurabhsandilya@gmail.com}}

\abstract{
  Belle II is an upcoming experiment at the next generation SuperKEKB
  $e^{+}e^{-}$ collider. It will record about 50 \invab of data mostly
  at the \Y4S resonance. The high statistics will allow precision
  measurements for rare decays, which can be then compared with the
  Standard Model predictions in order to search indirectly for new
  physics. We present the prospects for rare decay studies at Belle II.
}

\FullConference{9th International Workshop on the CKM Unitarity Triangle\\
		28  November - 3 December 2016\\
		Tata Institute for Fundamental Research (TIFR), Mumbai, India}

\begin{document}

\section{Introduction}
\label{sec:intro}
B-factories had a successful operational period with a total recorded sample of over 1.5 \invab of data corresponding to $1.2 \times 10^{9}$ B-meson pairs.
As a next generation B-factory, the Belle II experiment will search for new physics (NP) beyond the Standard Model (BSM).
The Belle II detector will be located at the upgraded SuperKEKB $e^{+}e^{-}$ asymmetric collider.
It will record most of its data at the \Y4S resonance, which gives a very clean sample of quantum correlated B-meson pairs.
Belle II is expected to increase the data-sample size by a factor of 50 compared to Belle. 
The SuperKEKB accelerator has been designed to give an instantaneous luminosity of $8 \times 10^{35}\rm{cm^{-2}s^{-1}}$, which is about 40 times larger than the previous KEKB accelerator.
The first data taking run for physics analyses is expected to begin in early 2018.
In this proceedings we present the prospects for rare decay studies at Belle II.

\section{Inclusive \bm{$B\to X_{s,d} \gamma$}}
\label{sec:btosgamma}
The $b \to (s,d) \gamma$ transition is a flavor changing neutral current (FCNC) process forbidden at tree level in the Standard Model (SM) and proceeds via a radiative `penguin' loop diagram.
These types of decays are sensitive to potential contributions from non-SM particles.
The NP scenarios can be investigated through a precise measurement of the branching fraction (BF) of the inclusive $B\to X_{s,d} \gamma$ decays.
The BF is predicted very precisely in the SM with ${\cal B}_{s\gamma } = (3.36 \pm 0.23) \times 10^{-4}$ and ${\cal B}_{d\gamma } = (1.73^{+0.12}_{-0.22}) \times 10^{-5}$, for $E_{\gamma} > 1.6 ~\gev$~\cite{misiak2015}.
Recently, Belle measured the BF for $B\to X_{s} \gamma$ process, using a fully inclusive method and the preliminary result (extrapolated for $E_{\gamma} > 1.6 ~\gev$) ${\cal B}_{s\gamma } = (3.12 \pm 0.10 {\rm (stat.)} \pm 0.19 {\rm (sys.)} \pm 0.08 {\rm (model)} \pm 0.04 {\rm (extrap.)} ) \times 10^{-4}$ is the world's most precise measurement, which is in agreement with the SM prediction as well as previous measurements~\cite{belle_xgamma}.
Evaluation of the constraint on BSM scenario depends crucially on both the central value and the uncertainties on the BF.
The above mentioned Belle result with a fully inclusive method has 7.3\% uncertainty and excludes a mass of the charged Higgs boson below $580 ~\gev$ at 95\% confidence level.

\begin{figure}[htp]
  \centering
  \includegraphics[width=0.45\textwidth]{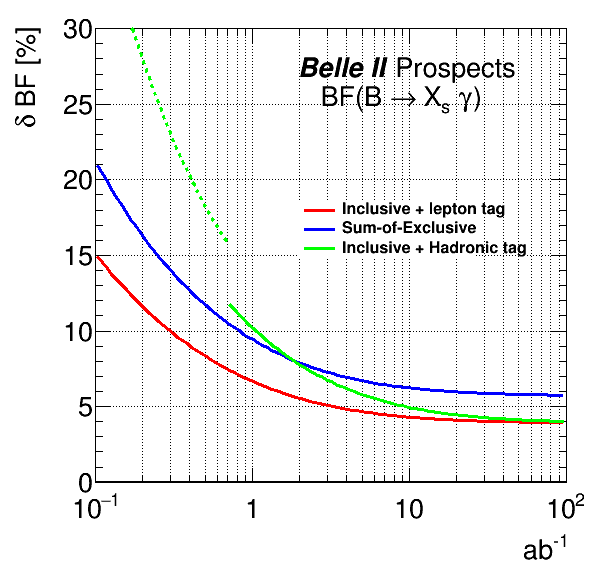}
  \includegraphics[width=0.45\textwidth]{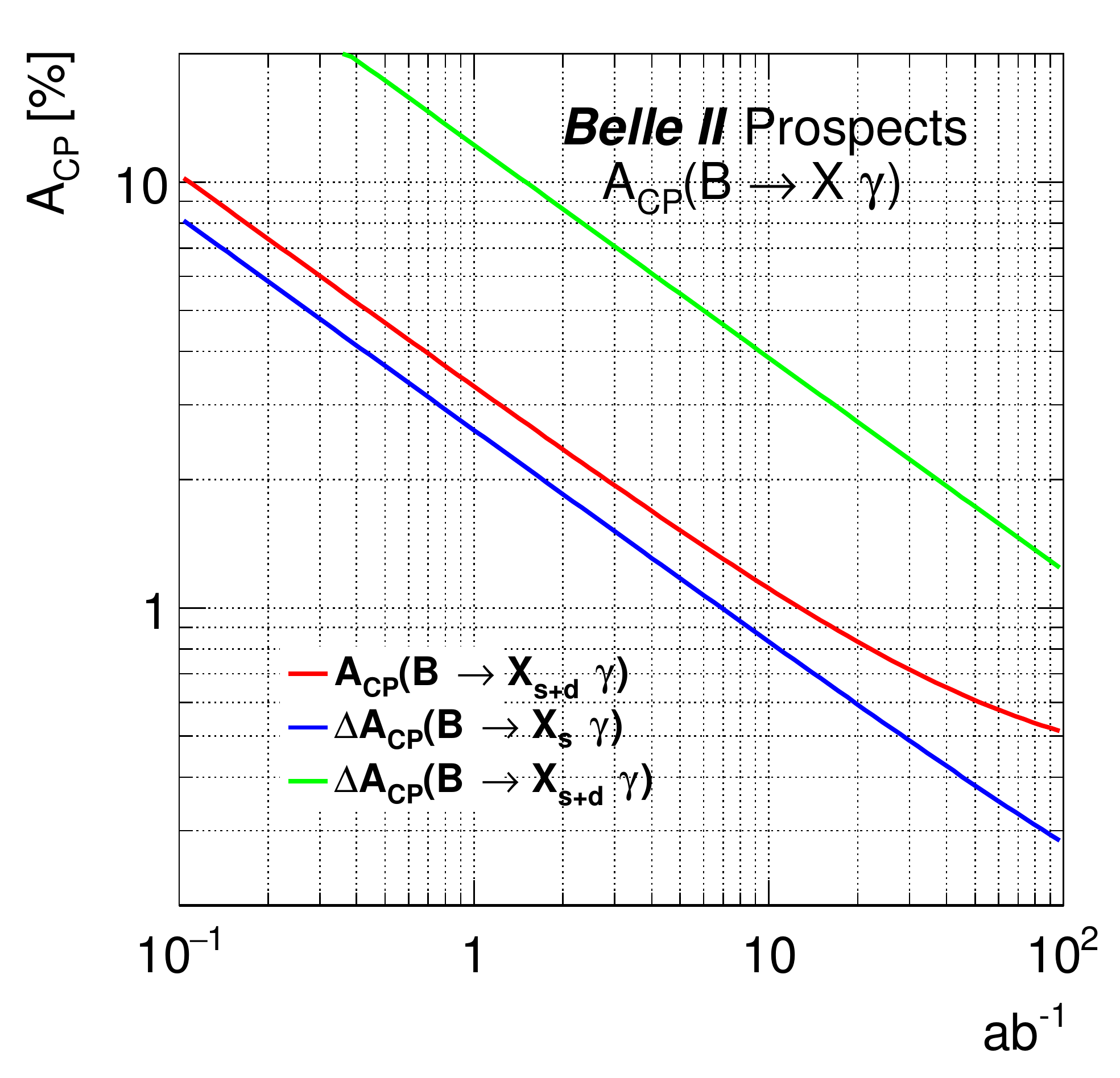}
  \caption{Belle II projections for uncertainities in inclusive ${\cal B}(B\to X_{s} \gamma)$ and $A_{CP} (B\to X \gamma)$.}
  \label{fig:proj1}
\end{figure}

Belle II is expected to reduce the systematic uncertainty in the measurement with its large data sample. Conservatively estimated, 3.9\% total error will be achievable with 50 \invab; see the projection plot in Figure~\ref{fig:proj1} (left).
This is comparable to the theory uncertainty due to non-perturbative effects (which is hard to reduce)~\cite{misiak2015}.
In addition, at Belle II it will be also possible to measure the BF with $E_{\gamma} > 1.6 ~\gev$, then there won't be any need for extrapolation to compare the results with theoretical predictions.  

In addition to the BFs, isospin and CP asymmetries in the decay rates are also sensitive to the BSM contributions.
The uncertainty on the isospin measurements can be improved at Belle II with more statistics.
The SM predicts quite different CP asymmetries for $B\to X_{s} \gamma$ and $B\to X_{d} \gamma$, however for the sum of these decays it is predicted to be very small (close to zero, due to the unitarity of the CKM matrix).
Further, the difference of $A_{CP} (B\to X_{s} \gamma)$ between charged and neutral B mesons ($\Delta A_{CP}$) is sensitive to phases in Wilson coefficients $C_{7}$ and $C_{8}$, which is zero.
So, if either  $A_{CP} (B\to X_{s+d} \gamma)$ or $\Delta A_{CP}$ is deviated from zero, it will be a clear NP signal~\cite{th_asym1,th_asym2}.
In asymmetry measurements, most of the systematic errors cancel out, so both will be still statistically dominated at Belle II with 50 \invab.
The uncertainties in $A_{CP} (B\to X_{s+d} \gamma)$ and $\Delta A_{CP}$ are projected to be $\pm 0.61 \%$ and  $\pm 0.37 \%$, respectively.
The Belle II projection plots for $A_{CP}$ and $\Delta A_{CP}$ are shown in Figure~\ref{fig:proj1} (right). 

\section{Exclusive \bm{$b\to s \gamma$} }
\label{sec:exbtosgamma}

\begin{figure}[htp]
  \centering
  \includegraphics[width=0.45\textwidth]{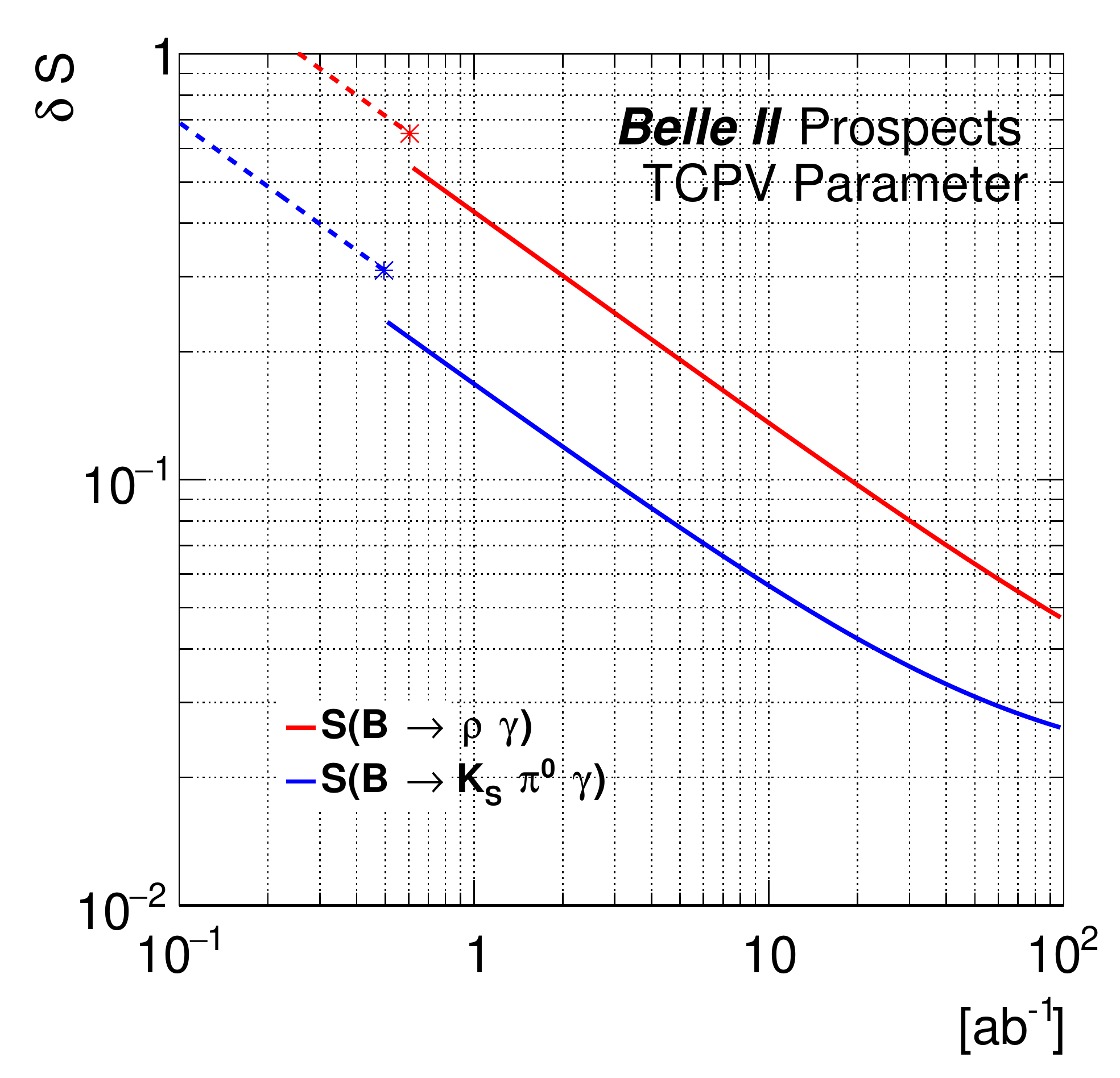}
  \includegraphics[width=0.45\textwidth]{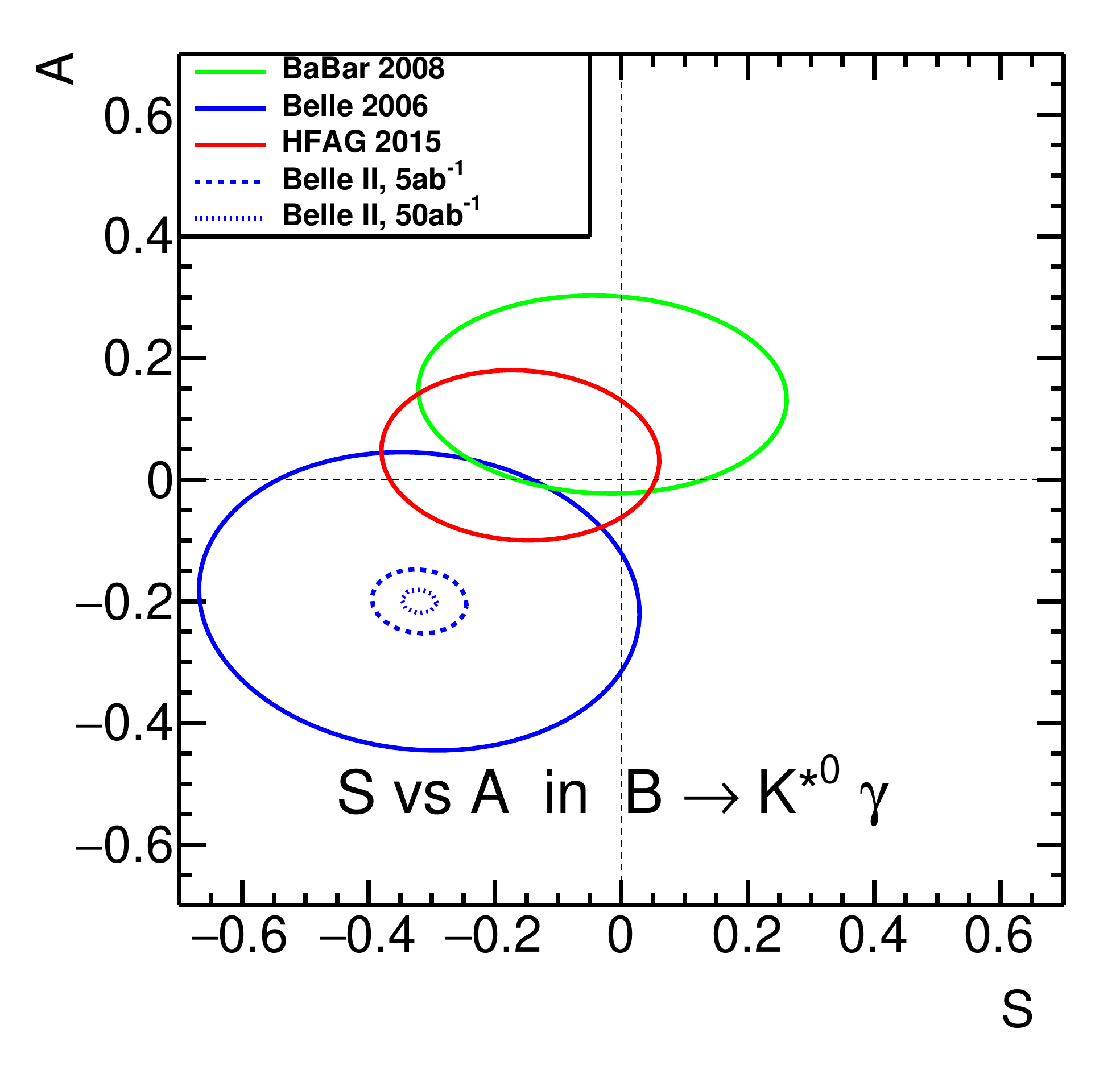}
  \caption{Belle II projections for uncertainities in inclusive ${\cal B}(B\to X_{s} \gamma)$ and $A_{CP} (B\to X \gamma)$.}
  \label{fig:proj2}
\end{figure}

Mixing-induced CP asymmetry in an exclusive $b\to s \gamma$ CP eigenstate mode such as $B\to K^{\star}[K_{s}^{0}\pi^{0}]\gamma$ is an excellent probe for a particular class of NP scenario~\cite{tdcpv1}.
In the SM, expected asymmetry is $S^{SM}_{K^{\star}[K_{s}^{0}\pi^{0}]\gamma} = -(2.3 \pm 1.6)\%$ and $S^{SM}_{\rho^{0}[\pi^{+}\pi^{-}]\gamma} = -(0.2 \pm 1.6)\%$~\cite{tdcpv2,tdcpv3,formfactor}.
As the two final states from $B^{0}$ and $\bar{B^{0}}$ decays have photons of different helicity (opposite helicity photon is suppressed by the $m_s$/$m_b$ factor) and therefore do not mix.
NP contribution with the right handed current would increase the fraction of right handed photon, and its interference with the SM can give large time dependent CP violation.
Studies of these asymmetries are thus considered to be one of the most promising methods to search for the BSM right-handed currents.
At Belle II, the vertex detector is larger than at Belle (11.5 cm radius in Belle II {\it cf.} 6 cm), which will give 30\% more $K_{s}^{0}$ with vertex hits.
And also, effective tagging efficiency is 13\% larger than at Belle (conservative estimation), hence significant improvement in the determination of $A_{CP}(t)$ in $B\to K^{\star}[K_{s}^{0}\pi^{0}]\gamma$ is expected.
The Belle II projection plot for uncertainty in $A_{CP}(t)$ is shown in Figure~\ref{fig:proj2} (left). With the Belle II projected uncertainty, the central value measured by Belle~\cite{acp_belle} would be $16\sigma$ deviation from zero, as shown in Figure~\ref{fig:proj2} (right).

\section{\bm{$b\to s \ell^{+} \ell^{-}$}}
\label{sec:btosll}
The $b\to s \ell^{+} \ell^{-}$ is also an FCNC process, which is sensitive to NP.
Owing to lepton universality in the SM, the ratio between the BFs of the electron mode to the muon mode is expected to be unity.
LHCb reported 2.6$\sigma$ deviation of $R_{K}$ (ratio of the BFs between $B^{+}\to K^{+}\mu^{+}\mu^{-}$ and $B^{+}\to K^{+}e^{+}e^{-}$) from the SM expectation for low $q^{2}$ region~\cite{LHCb:RK}.
Since the reconstruction of electron modes at Belle II is easier than at LHCb and have a comparable efficiency to that for muon modes.
The ratios $R_{K}$, $R_{K^{\star}}$ and $R_{X_{S}}$ can be measured precisely in Belle II for both low and high $q^{2}$ regions.
In ratio measurements most of the systematic error cancels out except for lepton identification, which is expected to be about 0.4\%.
The error will be mostly statistical dominant even with the Belle II data.

Measurement in the inclusive $B\to X_{s} \ell^{+} \ell^{-}$ channel is theoretically cleaner than the exclusive channels, especially for $q^{2}$ region below the charm resonances~\cite{xsll:th1,xsll:th2,xsll:th3,xsll:th4}.
The measurements for $B\to X_{s} \ell^{+} \ell^{-}$ decays in BaBar~\cite{xsll:babar1,xsll:babar2} and Belle~\cite{xsll:belle1} suffer from a sizeable experimental uncertainties.
Furthermore, these measurements are based on a sum over several exclusive states, which makes a direct comparison to the theoretical predictions difficult.
It is expected from Belle II to improve upon the present situation.
Belle has performed the first measurement of the forward-backward asymmetry ($A_{FB}$) in $B\to X_{s} \ell^{+} \ell^{-}$ with the sum of several exclusive modes~\cite{afb:belle}.
The $A_{FB}$ is found to be mostly consistent with the theoretical prediction~\cite{afb:th} with a mild tension in the low $q^{2}$ region.
The measurement of $A_{FB}$ can also be improved at Belle II, where the errors are projected to be in order of few \% with the 50 \invab data. 

\begin{figure}[htp]
  \centering
  \includegraphics[width=0.45\textwidth]{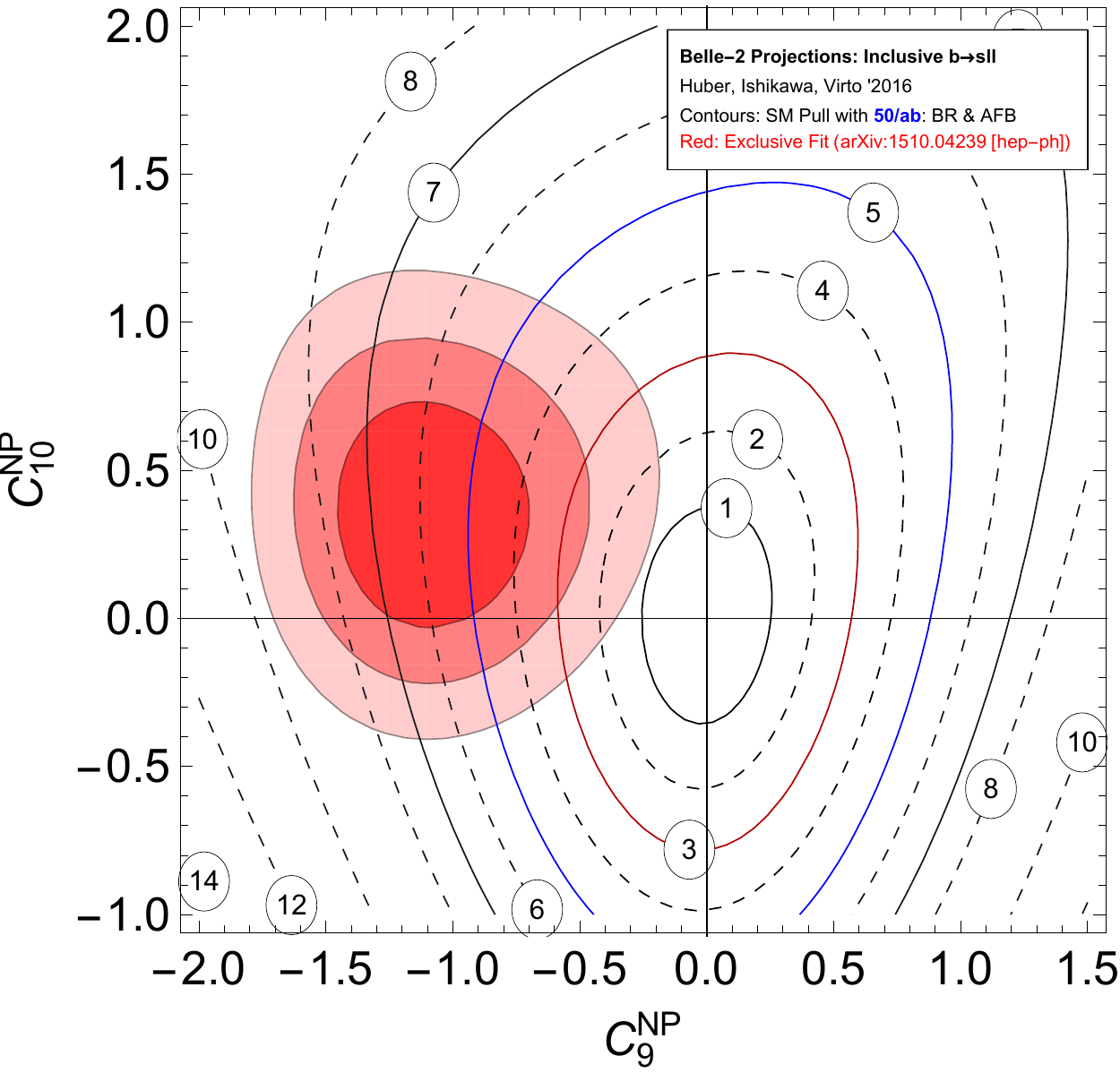}
   \caption{Potential of inclusive $b\to s \ell^{+} \ell^{-}$ measurements at Belle II  with current exclusive constraints~\cite{superplot}.}
  \label{fig:proj3}
\end{figure}

In order to provide a meaningful insight to the impact of BF and $A_{FB}$ measurements for $B\to X_{s} \ell^{+} \ell^{-}$ in Belle II, we look at the potential for model-independent constraints on the relevant Wilson-coefficients derived from inclusive measurements only.
Such projection is shown in Figure~\ref{fig:proj3}, considering the Wilson coefficients $C_{9}$ and $C_{10}$ as the ones potentially receiving relevant NP contributions~\cite{superplot}.

\section{\bm{$b\to s \nu \bar{\nu}$}}
\label{sec:btosnn}
The $b\to s \nu \bar{\nu}$ decays are the theoretically cleanest among the FCNC processes~\cite{bsnn:th}.
The BF of the decay $B\to K^{(\star)} \nu \bar{\nu} $ is mainly limited by $B\to K^{(\star)}$ form factors and relevant CKM matrix elements.
The SM predictions are available in Ref.~\cite{bsnn:th} and recently updated in Ref.~\cite{bsnn:th1}.
Belle very recently updated $b\to (s,d) \nu \bar{\nu}$ BF measurement with semileptonic tagging~\cite{bsnn:belle}.
These decays can be observed with Belle II, assuming the SM prediction holds.
Belle II will be able to provide a measurement with uncertainties of similar size as the current theoretical uncertainties. 

\section{Summary and Status}
\label{summary}
Belle II has a rich physics program; both complementary to, and competitive with, the LHCb experiment and energy frontier flavor physics programs. With the more powerful Belle II detector and higher luminosity machine SuperKEKB, we can search with high statistics for NP in radiative and electroweak penguin decays.
Accelerator commissioning (phase 1) was successful in June 2016.
Phase 2 without vertex detector will start in the end of 2017 and phase 3 with the complete detector is expected in the fall of 2018.
The Belle II detector is now mostly installed and currently being commissioned for the phase 2 running.


\begin{thebibliography}{99}

\bibitem{misiak2015}
  M. Misiak {\it et al.}, 
  Phys.\ Rev.\ Lett.\ {\bf 114}, 221801 (2015).

\bibitem{belle_xgamma}
  A. Abdesselam {\it et al.} (Belle Collaboration),
  arXiv:1608.02344.

\bibitem{th_asym1}
  T. Hurth, E. Lunghi, and W. Porod, 
  Nucl.\ Phys.\ {\bf B704}, 56 (2005).
  
\bibitem{th_asym2}
  M. Benzke, S.J. Lee, M. Neubert, and G. Paz,  
  Phys.\ Rev.\ Lett.\ {\bf 106}, 141801 (2011).

\bibitem{tdcpv1}
  D. Atwood, M. Gronau, and A. Soni, 
  Phys.\ Rev.\ Lett.\ {\bf 79}, 185 (1997).

\bibitem{tdcpv2}
  P. Ball, G.W. Jones, and R. Zwicky, 
  Phys.\ Rev.\ D {\bf 75}, 054004 (2007).

\bibitem{tdcpv3}
  P. Ball, and R. Zwicky, 
  Phys.\ Lett.\ B {\bf 642}, 478 (2006).

\bibitem{formfactor}
  A. Bharucha, D. M. Straub, and R. Zwicky,
  JHEP {\bf 08}, (2016) 098.

\bibitem{acp_belle}
  Y. Ushiroda {\it et al.} (Belle Collaboration),
  Phys.\ Rev.\ D {\bf 74}, 111104(R) (2006).

\bibitem{LHCb:RK}
  R. Aaij {\it et al.} (LHCb Collaboration),
  Phys.\ Rev.\ Lett.\ {\bf 113}, 151601 (2014).

\bibitem{xsll:th1}
  C. Bobeth, P. Gambino, M. Gorbahn, and U. Haisch,
  JHEP {\bf 04}, 071 (2004).
  
\bibitem{xsll:th2}
  T. Huber, E. Lunghi, M. Misiak, and D. Wyler,
  Nucl.\ Phys.\ {\bf B740}, 105 (2006).
  
\bibitem{xsll:th3}
  T. Huber, T. Hurth, and E. Lunghi,
  Nucl.\ Phys.\ {\bf B802}, 40 (2008).
  
\bibitem{xsll:th4}
  T. Huber, T. Hurth, and E. Lunghi,
  JHEP {\bf 06}, 176 (2015).

\bibitem{xsll:babar1}
  B. Aubert {\it et al.} (BaBar Collaboration),
  Phys.\ Rev.\ Lett.\ {\bf 93}, 081802 (2004).
  
\bibitem{xsll:babar2}
  J. P. Lees {\it et al.} (BaBar Collaboration),
  Phys.\ Rev.\ Lett.\ {\bf 112}, 211802 (2014).
  
\bibitem{xsll:belle1}
  M. Iwasaki {\it et al.} (Belle Collaboration),
  Phys.\ Rev.\  D {\bf 72}, 092005 (2005).

\bibitem{afb:belle}
  Y. Sato {\it et al.} (Belle Collaboration),
  Phys.\ Rev.\  D {\bf 93}, 032008 (2016).

\bibitem{afb:th}
  S. Fukae, C. S. Kim, T. Morozumi, and T. Yoshikawa,
  Phys.\ Rev.\ D {\bf 59}, 074013 (1999).

\bibitem{superplot}
  T. Huber, A. Ishikawa, and J. Virto (private communication),
  to be published in B2TiP report.

\bibitem{bsnn:th}
  A. J. Buras, J. Girrbach-Noe, C. Niehoff, and D. M. Straub,
  JHEP {\bf 02}, 184 (2015).

\bibitem{bsnn:th1}
  D. M. Straub,  BELLE2-MEMO-2016-007 (2016).

\bibitem{bsnn:belle}
  J. Grygier {\it et al.} (Belle Collaboration),
  arXiv:1702.03224
  
\end{thebibliography}
\end{document}